# T$_2$* and Susceptibility Mapping as Indicators of Placental Health

Amy Turnbull[1], George Hutchinson[1], Louise Dewick[2], Ruizhe Li[3], Chris Bradley[1,4], Lopa Leach[5], Dimitrios Amantis[5], Xin Chen[3], Grazziela Figueredo[3] Kate F Walker[2] and Penny Gowland[1,4]

[1]Sir Peter Mansfield Imaging Centre, University of Nottingham, Nottingham, United Kingdom

[2]Centre for Perinatal Research, School of Medicine, University of Nottingham, Nottingham, United Kingdom

[3]School of Computer Science, University of Nottingham, Nottingham, United Kingdom

[4]National Institute for Health Research, Biomedical Research Centre, Hospital NHS Trust and University of Nottingham, Nottingham, United Kingdom

[5] School of Life Sciences, University of Nottingham, Nottingham, United Kingdom

# Abstract

Objective(s)

T2* and susceptibility (χ) MRI mapping provide complimentary measures of the haemodynamic environment in the placenta. The aims of this work were to use these simultaneously obtained measures to investigate the role of oxygen distribution on the well-established reduction of T2* with gestational age found in healthy pregnancies and explore differences in both measures in compromised placentas.

Methods

T2* and χ were measured simultaneously from a double echo, echo planar scan of the whole placenta, across a range of gestational ages and pregnancy complications. Regional variations across the placenta were investigated.

Results

Whole placental mean T2* was more correlated with standard deviation of χ than mean χ indicating it is more driven by increasing local inhomogeneities rather than bulk deoxygenation with healthy gestation. Compromised placentas also showed increased standard deviation of χ as well as lower mean T2* suggesting flow/uptake mismatch and reduced oxygenation.

Regionally, the susceptibility was lowest (most oxygenated) and least variable in the central region of the placenta indicating good mixing and refreshment of blood in this area. The susceptibility was highest (most deoxygenated) and most variable at the fetal side, suggesting less effective perfusion in this region. Compromised cases showed the greatest difference on the fetal side for both mean and standard deviation of χ. T2* was lowest at the fetal side for healthy and compromised cases but the maternal and central regions better distinguished between the two groups.

Conclusion(s)

T2* and susceptibility can be mapped simultaneously from a single MRI scan and provide complimentary information about the function of the placenta across healthy gestational development, and as a potential indicator of placental compromise.

# Introduction

The placenta enables exchange of oxygen, nutrients, and waste between maternal and fetal circulations. Exchange occurs between the fetal circulation, confined within villi which are bathed in maternal blood within the intervillous space. Disruption of this exchange can result in fetal growth restriction (FGR), and placental dysfunction contributes to 35% of stillbirths in the UK[1].

Placental T2* is recognised as an indicator of placental function, decreasing with gestational age (GA) and typically reduced in conditions associated with placental compromise, including pre-eclampsia and FGR[2–22]. This reduction is often attributed to changes in blood oxygenation altering the magnetic susceptibility of haemoglobin. However, while T2* is sensitive to bulk oxygenation, it is also strongly influenced by local oxygenation heterogeneity, shown by the Blood Oxygen Level Dependent (BOLD) effect used in functional MRI[23]. Additionally, T2* is affected by blood flow, blood/tissue volume fraction, and susceptibility-altering inclusions such as haematomas or calcifications. These factors often change with GA and in compromised placentas[24–28].

Quantitative susceptibility mapping (QSM) offers more specific sensitivity to blood oxygenation and may therefore provide complementary information. Although QSM is also influenced by tissue components such as calcification and haematoma, the high blood volume fraction in the placenta means blood oxygenation is likely to dominate. Importantly, QSM can be derived from the same gradient echo acquisitions used for T2* mapping but provides a physically independent measure.

Placental QSM has previously been shown to be feasible at 1.5 and 3T. Mean whole placental susceptibility[18,21,29] showed little gestational age dependency, however the standard deviation of susceptibility has been reported to increase with gestational age[29] and previous work by our group showed larger variation (interquartile range) across the placenta in pre-eclampsia[21]. These studies also reported T2* but made no direct comparison between the measures.

We recruited participants to a longitudinal study in the third trimester, some with a range of pregnancy risk factors. The aim was to directly compare magnetic susceptibility and T2* to better understand the mechanisms underlying T2* changes, and to characterise how these metrics evolve across gestation in both healthy and compromised placentas, and investigate regional variation within the placenta.

## Method

### Recruitment

The datasets reported here were acquired as part of a larger protocol (Phases A and B of SWIRL: Stillbirth When Is Risk Low?). Ethics approval was obtained from the East Midlands-Leicester Central Research Ethics Committee (23/EM/0052). Participants were recruited from Nottingham University Hospitals NHS Trust and written, informed consent was obtained from each participant.

78 pregnant participants with gestational ages from 29-41 weeks attended for between one and five visits. Inclusion criteria were women with singleton pregnancies, aged 18-50 years old, able to give informed consent, body mass index (BMI) ≤45 kg/m$^2$, no known fetal congenital anomalies and no contraindications to MRI. Most participants had a blood test for placental-like growth factor (PLGF)/soluble FMS-like tyrosine kinase-1 (SFlt-1) levels during their first visit (N = 66) and clinical information was recorded following birth for all participants, including a histological examination of their placenta where possible (N = 60).

This work was part of a study aimed at predicting fetal outcome and hence a very varied compromised group was recruited. However for this analysis, participants have been categorised as healthy controls (HC), mild compromise (mild-moderate pre-eclampsia, gestational hypertension, abnormal PLGF/SflT levels, mild abnormalities on histology), and severe compromise (severe pre-eclampsia, FGR, perinatal hypoxia at birth, severe vascular malperfusion on histology, placental abruption). Data is shown for each group separately, but statistical analyses simply compared healthy to compromised (including both mild and severe groups together) groups to increase statistical power.

### MRI Protocol

Participants were scanned in a lateral tilt to reduce vena cava compression. Images were acquired with a 3T Philips Ingenia MRI scanner using a d-stream posterior bed coil and anterior body coil. The anterior coil was placed as close as possible to the placenta if the location was known prior to scanning and repositioned after initial localiser scans if required to improve sensitivity over the placenta. After acquiring an initial half-fourier acquisition single-shot turbo spin echo (HASTE) dataset to visualise locations clearly, a set of double echo, echo planar imaging (EPI) scans covering the whole placenta were acquired in 10 seconds with TE=14/38ms, 25 fixed transverse slices, slice gap 5mm, voxel size 3x3.2x3, FOV=400x400x195mm (FOV=420x420x195mm for N = 6 larger participants), bandwidth per pixel 45.6 Hz, no SENSE. Volume shimming was applied to a region encompassing the whole placenta, this was angled where required. FOV increase and fat shift direction was chosen to avoid clipping the body and introducing artefacts in the placenta. Other scans were then acquired but are not reported here.

Data Analysis

Images were segmented using an nnU-Net model[30] trained on 169 volumes from 39 individuals of a similar T2*-weighted scan and manually edited where necessary. The double echo magnitude data was used to calculate T2*. Phase images from the first echo were unwrapped, background phase was removed and QSM maps (χ maps) were calculated using STI Suite version 3, MATLAB 2023b.

The mean and standard deviation of T2* and χ (T2*$_{mean}$, T2*$_{SD}$, χ$_{mean}$, χ$_{SD}$) were calculated across the whole placenta.

To investigate the changes in compromised pregnancy without the confounding effects of variations with GA, healthy controls were used to estimate a linear mixed effect model with random intercepts, centred at a GA of 28 weeks of T2*$_{mean}$, T2*$_{SD}$, χ$_{mean}$ and χ$_{SD}$ against GA, and subsequently Z scores of the deviation of all data points from this model were calculated ($Z_{GA}$).

Visits were categorised as <34 weeks or ≥ 34 weeks GA to probe the effect of early and late onset compromise, since pregnancies with early onset compromise were often delivered before 34 weeks and those with late onset compromise were often only recruited after 34 weeks. For cases where participants had more than one visit within the range, the visit closest to the centre of the GA range was chosen (31.5 and 37.5 weeks). Statistical comparisons between healthy and compromised groups were calculated using a t-test. Bonferroni correction was used to calculate the adjusted statistical significance threshold for the four measurements at both GA ranges considered (T2*$_{mean}$, T2*$_{SD}$, χ$_{mean}$, χ$_{SD}$), and so a threshold of $p = \frac{0.05}{8} = 0.00625$ was used to indicate significance.

To further investigate the variation of T2* and χ across the placenta, the placental mask was divided into maternal, fetal and central regions. For each voxel the shortest distance to the basal plate and chorionic plate was calculated, and then normalised to the maximum thickness of the placenta. The maternal region was defined as within 25% of the maximum thickness from the basal plate, fetal region as within 25% from the chorionic plate and central as further than 25% thickness from both. These regional masks were applied to the T2* and χ maps to calculate T2*$_{mean}$, T2*$_{SD}$, χ$_{mean}$ and χ$_{SD}$ in those regions. Variations between the regions for the first visit of each participant were compared using an ANOVA to test for significance and Tukey HSD for pairwise comparisons for healthy controls and compromised placentas.

As for the full placenta, linear mixed effects models were estimated for regional values of T2*$_{mean}$, T2*$_{SD}$, χ$_{mean}$ and χ$_{SD}$ across GA and Z$_{GA}$ scores calculated for each visit. Again, these were then categorised as <34 or ≥ 34 weeks GA and the compromised cases were compared to healthy controls. Bonferroni corrected adjusted statistical significance threshold for regional comparisons was $p = \frac{0.05}{24} = 0.0021$.

## Results

Table S1 summarises the clinical data for all participants. 51 participants were categorised as HC (PLGF/SflT levels not available in 12 of these participants who had been recruited as low risk according to the Saving Babies' Lives Care Bundle Version 2 (NHS England))[31]. The HC group also included one case of perinatal hypoxia due to impacted fetal head at caesarean birth with no pre- or postnatal indications of placental compromise.

26 participants were categorised as having compromised placentas: 13 of these were mild compromise and 13 were severe.

One participant was removed from analysis due to complications which were not directly related to placental function (trisomy 21), but Z$_{GA}$ scores have been included in the tables and indicate no particular change in these parameters compared to healthy controls.

Figure 1 shows typical example T2* and χ maps of a single slice from a gestational-age matched healthy and compromised placenta. These illustrate that in compromised pregnancies the placenta is generally smaller, with a shorter T2* and more variable susceptibility.

Figure 2a shows the variation of T2*$_{mean}$, T2*$_{SD}$, χ$_{mean}$ and χ$_{SD}$ with GA for healthy controls with linear mixed effect model and 95$^{th}$ centile CI for the healthy population superimposed; T2*$_{mean}$ decreased with gestational age as expected ($p < 10^{-10}$), as did T2*$_{SD}$ although to a lesser extent ($p = 6.0 \times 10^{-6}$). χ$_{mean}$ showed no apparent change across gestation (p = 0.25), however χ$_{SD}$ increased with GA ($p < 10^{-10}$).

Figure 2b shows the data for the compromised group with the HC linear model from Figure 2a superimposed, showing that compromise cases had lower T2*$_{mean}$ and T2*$_{SD}$ and higher χ$_{mean}$ and χ$_{SD}$ across the gestational age range.

Figure 2c plots the Z$_{GA}$ score for early and late onset compromise separately (before and after 34 weeks GA). T2*$_{mean}$ was significantly lower in mild and severe compromise below 34 weeks gestation (p = 0.000046) and beyond 34 weeks (p = 0.00047). T2*$_{SD}$ was also lower at both GA ranges, however this was only significant before 34 weeks (p = 0.0042 and p = 0.014 respectively). χ$_{mean}$ appeared higher in the compromised groups but was not significant at either GA range (p = 0.073 and p = 0.011 respectively). χ$_{SD}$ was significantly higher in compromised cases < 34 weeks but not > 34 weeks (p = 0.000063 and 0.011). Z$_{GA}$ scores for all visits are shown in Table S1. No particular trends can be seen between mild and severe compromise, except for χ$_{SD}$.

Figure 3 replots the Z$_{GA}$ scores against birth weight centile, showing that for all measures the largest change in Z scores in the compromised cases occur for pregnancies that resulted in low birth weights.

Spatial variation of χ across the placenta is shown in Figure 4 with statistical comparisons summarised in Table 1. In HC, T2*$_{mean}$ was lowest at the fetal side (p = 9.1x10$^{-6}$ and 3.1x10$^{-6}$ compared to maternal and central respectively) with no difference between the maternal and central regions. T2*$_{SD}$ was consistent across the placenta. χ$_{mean}$ was significantly higher (more deoxygenated) at the fetal side ($p < 10^{-10}$ compared to both maternal and central regions). χ$_{SD}$ was lower in the central region than the maternal or fetal regions ($p = 0.00039$ and $<10^{-10}$ respectively).

Compromised cases followed similar profiles in χ and T2*, although the changes were only significant for χ. Comparing compromised groups to healthy controls using $Z_{GA}$ (Table 2), the difference in T2*$_{mean}$ between HC and compromised cases were most apparent on the maternal side at both GA ranges, whereas χ$_{SD}$ was more different at fetal side.

Figure 5 plots the relationships between R2*$_{mean}$ (1/T2*) and R2*$_{SD}$, χ$_{mean}$ and χ$_{SD}$ (with R2* plotted instead of T2* since this inverse parameter is expected to depend more directly on absolute and local variations of oxygenation). R2*$_{mean}$ is most correlated with R2*$_{SD}$ ($R^2 = 0.88$). R2*$_{mean}$ is more correlated with χ$_{SD}$ than χ$_{mean}$ ($R^2 = 0.52$ and $0.054$ respectively for HC).

## Discussion

This study investigated rapid, simultaneous measurement of T2* and magnetic susceptibility (χ, a marker of blood oxygenation) in healthy and compromised placentas. It showed that T2* and the standard deviation of susceptibility are both changed in compromised placentas and has also provided new information on variations in these measures across the placenta. It offers new insights into the physiological origins of these changes in T2*.

These findings confirm and extend previous work suggesting that placental T2* reflects changes in placental function across GA and in compromised pregnancies[2–22]. Prior studies compared T2* with diffusion and blood flow[8,16,20,21], placental morphology[5,6,9,16] or association with fetal organ measurements[3,6,7,17,20]. However, relatively little work has explored spatial variations in T2*[6,9,20], and the mechanisms underlying its reduction with GA and placental compromise remain poorly understood. Despite its greater specificity to placental oxygenation, susceptibility mapping has seen limited application[18,21,29].

<u>Healthy pregnancies</u>

In healthy pregnancies, T2* decreased with GA, consistent with previous reports [2,3,5–7,9,11,13,14,16–18,20,22], whereas mean susceptibility (χ$_{mean}$) reflecting average oxygenation did not change. Although fetal growth might be expected to increase oxygen demand and reduce placental oxygenation, the placenta grows and maternal blood supply increases across gestation to meet fetal demand[32]. Prolonged net deoxygenation would be highly detrimental and not consistent with healthy pregnancy.

Nonetheless, oxygenation variability (χ$_{SD}$) did increase with GA, supporting a previous suggestion that greater oxygen extraction later in gestation might produce local flow/oxygen mismatches[33]. T2*$_{mean}$ correlated more strongly with χ$_{SD}$ than χ$_{mean}$ in healthy pregnancies, suggesting that the reduction in T2* with GA is probably driven by increasing heterogeneity of oxygenation rather than changes in bulk oxygenation. This is consistent with the well-known sensitivity of T2* to susceptibility variations (the BOLD effect[23]).

However, regional analysis revealed variation in χ$_{mean}$ across the placenta, with the fetal side being more deoxygenated. *In utero*, placental volume is dominated by maternal blood entering the intervillous space via spiral arteries in the basal plate and draining through venous outlets.

Ideally, perfusion would be uniform, but this study suggests relatively reduced oxygenation at the fetal side[26,34], where villous density is likely lower. *In vivo* gadolinium enhanced imaging has demonstrated long maternal blood transit times across the placenta, implying that oxygen may be partially depleted by the time blood reaches the fetal side[34]. Structural features on the fetal side, including stem villi and parallel fetal arteries and veins, may also influence voxel-averaged oxygenation.

The central placental region showed the most homogeneous oxygenation (lowest $\chi_{SD}$, lowest $\chi_{mean}$), suggesting effective mixing in an area of high villous density.

Placental compromise

Compromised placentas had reduced $T2^*_{mean}$ and increased oxygenation heterogeneity (higher $\chi_{SD}$), with a non-significant trend toward lower net oxygenation (higher $\chi_{mean}$). The modest change in net oxygenation is unsurprising since all pregnancies resulted in live births and it is unlikely a fetus could survive major sustained reductions in bulk oxygenation. The increased oxygenation heterogeneity probably reflects local flow/oxygen mismatches associated with abnormal perfusion and villous structure in conditions such as pre-eclampsia and FGR[27].

Profiles of $\chi_{mean}$ and $\chi_{SD}$ across the placenta in Figure 4 were broadly similar in healthy and compromised groups, both showing increased oxygenation and homogeneity centrally. However, compromised placentas exhibited greater deoxygenation and heterogeneity at the fetal side, consistent with impaired perfusion. Unlike in healthy cases, compromised $T2^*_{mean}$ did not decrease from the central region to the fetal side, possibly because central $T2^*_{mean}$ values were already markedly shortened.

$T2^*_{mean}$ differed most between healthy and compromised cases on the maternal side. This suggests these changes may reflect altered patterns of flow into or out of the intervillous space. Focusing analysis to the maternal region may enhance sensitivity in future $T2^*$-based assessments.

Technical issues, limitations and future work

$T2^*$ and $\chi$ were derived simultaneously from a gradient-echo acquisition, providing physically independent measurements: $T2^*$ from magnitude data and $\chi$ from phase data (the signal to noise ratio is inversely related between the magnitude and phase data, with a more complex relationship between the derived maps). The multi-echo acquisition reduced motion effects in $T2^*$. Future work could explore combined metrics to improve outcome group separation or condition-specific features.

The main limitation is the relatively small and heterogeneous compromised cohort. More tightly defined groups (e.g. pre-eclampsia only) may demonstrate clearer differences and improve understanding of condition-specific effects on oxygen distribution. Larger longitudinal studies could assess whether $T2^*/\chi$ mapping serves as a marker of evolving placental function and fetal wellbeing, potentially informing intervention decisions.

An assumption of this work is that magnetic susceptibility is a direct marker of blood oxygenation. However, susceptibility also depends on maternal and fetal blood volume fractions, other tissue compartments, haematocrit levels, and differences between adult and fetal haemoglobin susceptibility[35], which will complicate its use in absolute oxygenation measures.

The regional segmentation is a simplification of the complex perfusion patterns within the placenta and may be affected by variations in placental shape, with morphological changes often seen in compromised placentas[36]. These effects are minimised by considering relative placental thickness but neglects any intra-cotyledon interactions.

Finally, uterine and placental contractions are known to affect T2* and are likely to influence susceptibility[21,37–40]. Contractions were not explicitly accounted for in this study, as the aim was to evaluate a simple, rapid approach. Given that contractions may last up to 10 minutes[41], repeating the T2*/χ acquisition multiple times and selecting the highest T2* dataset could reduce contraction-related variability.

## Conclusion

T2* and QSM can be mapped simultaneously in the placenta using a rapid double-echo EPI acquisition. These complementary measures both develop with GA and are altered in compromised pregnancy, providing new insight into placental function. This approach has potential clinical relevance in managing at-risk pregnancies, but larger scale trials are required to evaluate its predictive value.

## Acknowledgements

This study was funded by the Wellcome Leap In Utero program. PhD funding for AT from EPSRC (EP/W524402/1). Publication funded by the Wellcome Trust.

## Collaborators

Nia Wyn Jones - Centre for Perinatal Research, School of Medicine, University of Nottingham, Nottingham, United Kingdom.

Cesar Peres - Centre for Perinatal Research, School of Medicine, University of Nottingham, Nottingham, United Kingdom.

Craig Platt - Centre for Perinatal Research, School of Medicine, University of Nottingham, Nottingham, United Kingdom.

## Data Availability

The data that support the findings of this study are available from the corresponding author upon reasonable request.

## Tables and Figures

| Regions compared | Maternal v. central | | Maternal v. fetal | | Central v. fetal | |
|---|---|---|---|---|---|---|
| Outcome group | HC | Comp | HC | Comp | HC | Comp |
| T2* mean | 0.98 | 0.98 | $9.1 \times 10^{-6}$ | 0.090 | $3.1 \times 10^{-6}$ | 0.059 |
| T2* SD | 0.89 | 0.97 | 0.60 | 0.82 | 0.32 | 0.93 |
| X mean | 0.0092 | $4.3 \times 10^{-4}$ | $< 10^{-10}$ | $< 10^{-10}$ | $< 10^{-10}$ | $< 10^{-10}$ |
| X SD | $3.9 \times 10^{-4}$ | $3.4 \times 10^{-4}$ | 0.0027 | 0.0066 | $< 10^{-10}$ | $< 10^{-10}$ |

Table 1: Results of ANOVA and Tukey HSD analysis of $T2^*_{mean}$, $T2^*_{SD}$, $\chi_{mean}$ and $\chi_{SD}$ variation across the maternal, central and fetal regions of the placenta for healthy controls (HC) and compromised cases (Comp). Uncorrected p values shown. Shading indicates values were significant at the Bonferroni corrected significance threshold.

| Metric (< 34 weeks) | Maternal | Central | Fetal |
|---|---|---|---|
| T2* mean | 6.3x10$^{-6}$ | 1.3x10$^{-4}$ | 2.4x10$^{-4}$ |
| T2* SD | 0.0021 | 0.010 | 0.013 |
| X mean | 0.38 | 0.19 | 0.0010 |
| X SD | 5.5x10$^{-4}$ | 0.034 | 7.5x10$^{-6}$ |
| Metric (≥ 34 weeks) | Maternal | Central | Fetal |
| T2* mean | 1.4x10$^{-5}$ | 1.3x10$^{-5}$ | 2.4x10$^{-4}$ |
| T2* SD | 2.3x10$^{-4}$ | 1.1x10$^{-4}$ | 0.011 |
| X mean | 0.069 | 0.11 | 0.020 |
| X SD | 0.019 | 0.093 | 0.008 |

Table 2: Comparison of $Z_{GA}$ scores between healthy controls and compromised cases for regional measurements of mean and standard deviation of T2* and magnetic susceptibility. Shown for < 34 weeks and > 34 weeks GA ranges. Shading indicates values were significant at the Bonferroni corrected significance threshold.

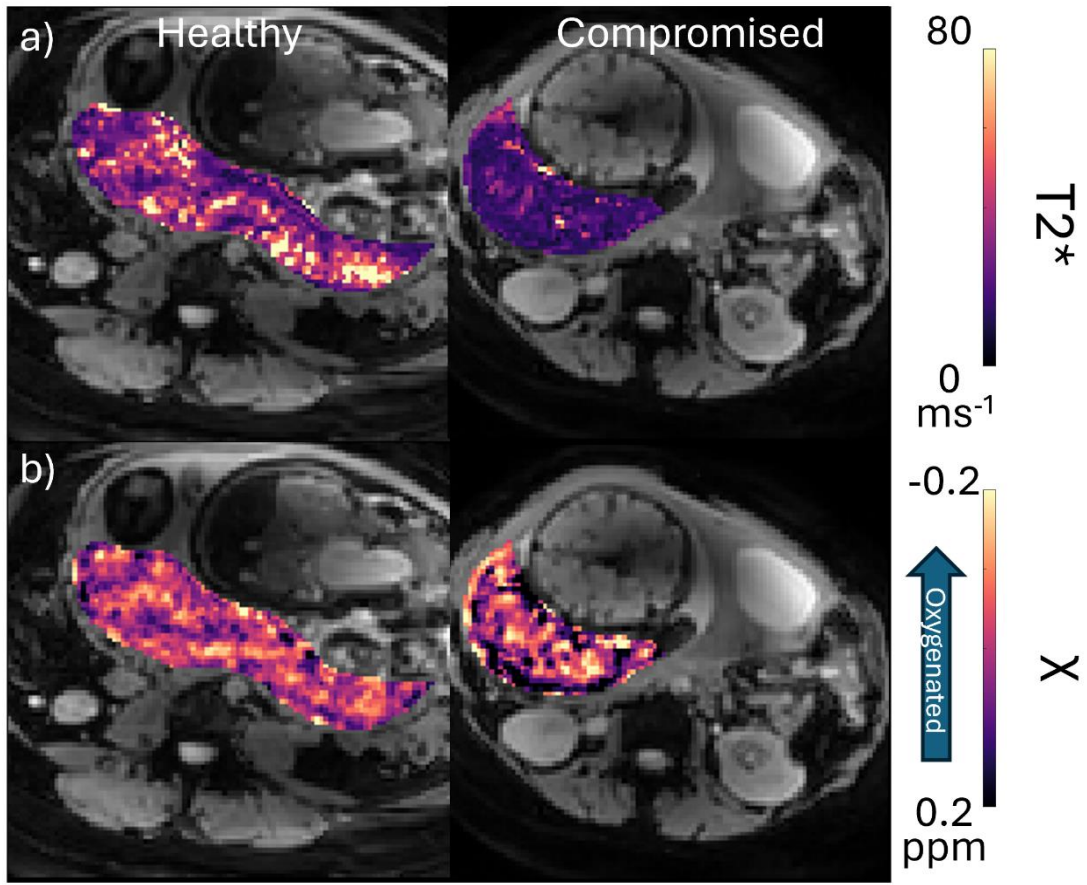

Figure 1: Example a) T2* and b) χ maps for a healthy and compromised case at similar gestational age, 36+5 and 36+2 respectively (weeks+days).

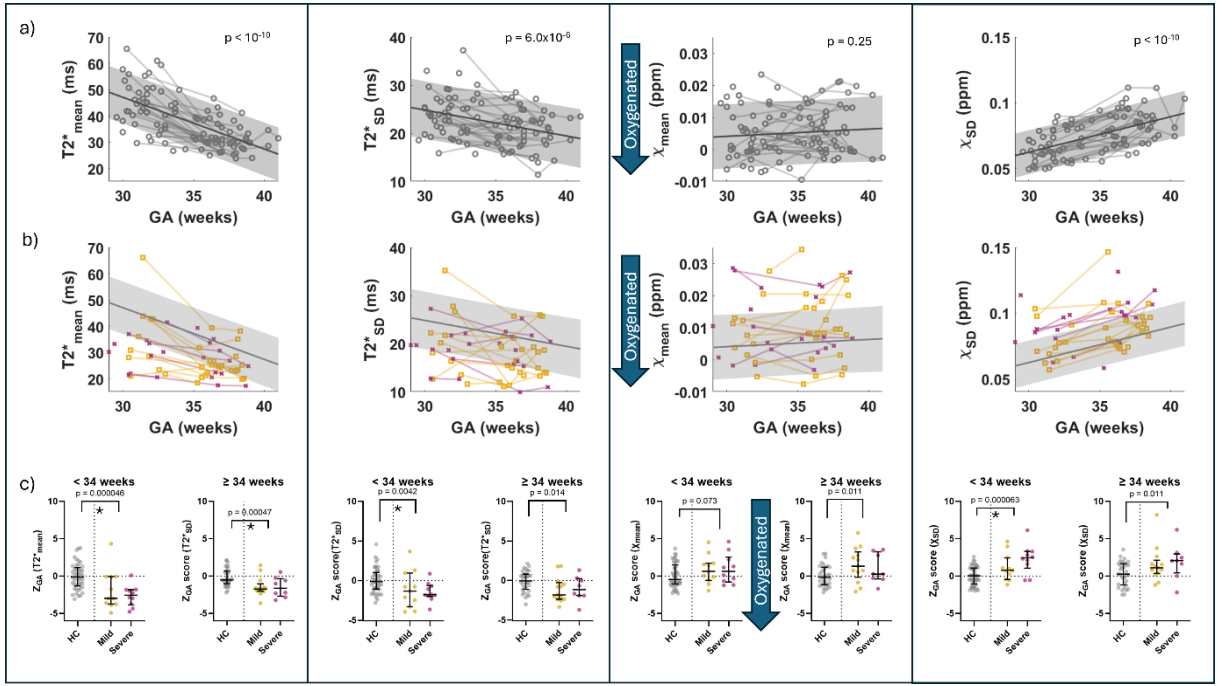

Figure 2: Change in T2*$_{mean}$, T2*$_{SD}$, χ$_{mean}$ and χ$_{SD}$ across gestational age, a) shows the linear mixed effects fit of healthy controls with 95% confidence interval plotted, b) shows the fit and confidence interval with the mildly and severely compromised cases, c) shows the z scores from the linear mixed effect model ($Z_{GA}$) for visits before 34 weeks and after 34 weeks GA for healthy controls and compromise groups. Significant differences between healthy controls and all placental compromise are marked with *, compared using a t-test and significance threshold Bonferroni corrected.

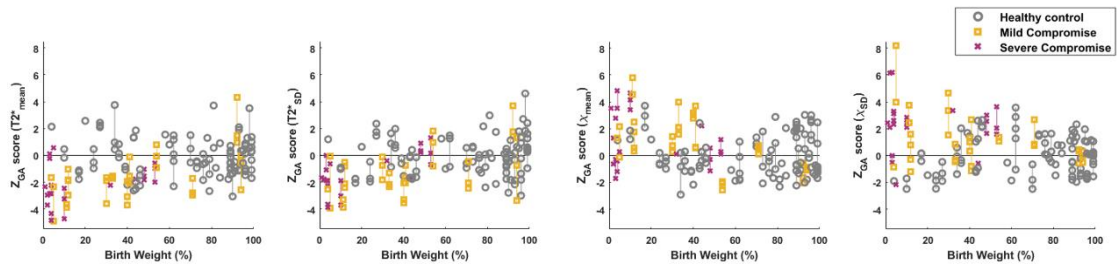

Figure 3: a) mean and standard deviation of T2* and susceptibility across gestational age for all outcome groups, longitudinal visits from the same participant are joined. The z scores for each visit calculated from the evolution of the healthy controls against b) gestational age and c) birth weight centile.

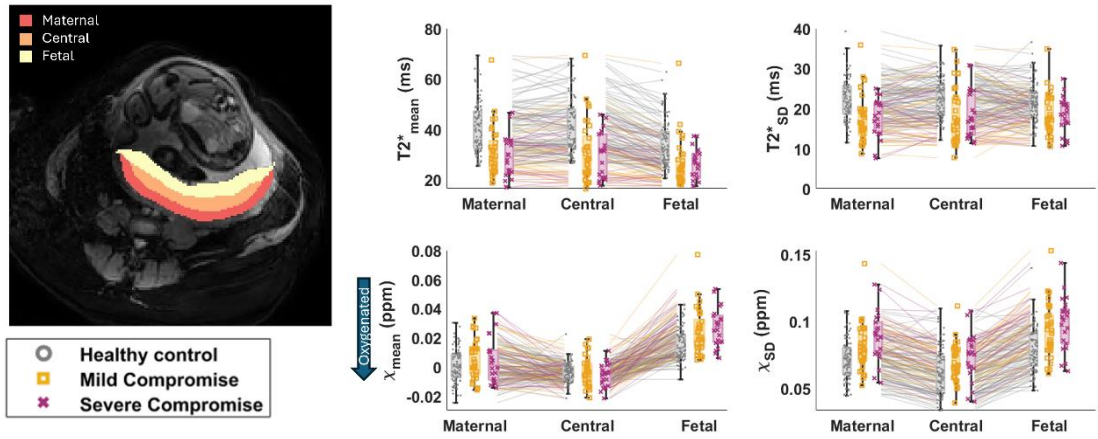

Figure 4: Spatial variation of T2* and susceptibility across the placenta, p values shown for comparisons between placental regions in healthy controls.

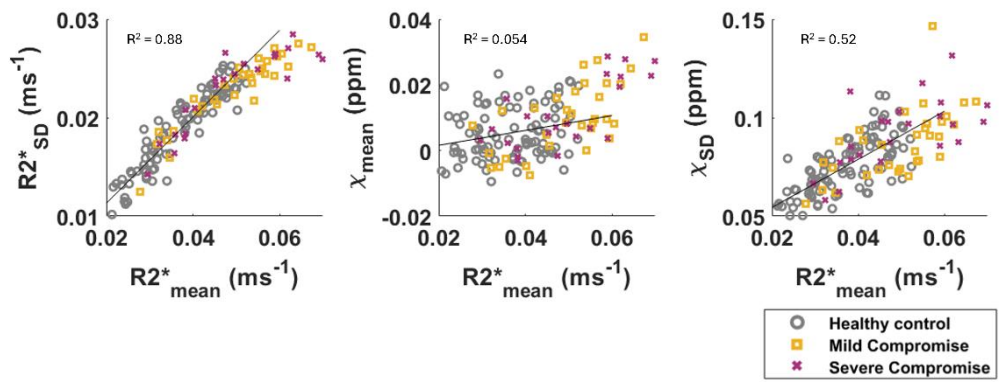

Figure 5: Relationship between mean R2* (1/T2*) and a) standard deviation of R2*, b) mean susceptibility, c) standard deviation of susceptibility. Including all visits where participants had multiple visits and not GA differentiated.